\documentclass[twocolumn, showpacs,superscriptaddress]{revtex4}
\usepackage{amsfonts}
\usepackage{amssymb}
\usepackage{graphicx}
\usepackage{dcolumn}
\usepackage{bm}
\usepackage[vcentermath]{youngtab}
\usepackage{amsmath}
\usepackage[title,titletoc,header]{appendix}

\begin{document}

\title{Entanglement dynamics of the ultra-strong coupling three-qubit Dicke model}
\author{Lijun Mao}
\affiliation{Institute of Theoretical Physics, Shanxi University, Taiyuan 030006, P. R.
China}
\author{Yanxia Liu}
\affiliation{Institute of Theoretical Physics, Shanxi University, Taiyuan 030006, P. R.
China}
\author{Yunbo Zhang}
\email{ybzhang@sxu.edu.cn}
\affiliation{Institute of Theoretical Physics, Shanxi University, Taiyuan 030006, P. R.
China}

\begin{abstract}
We give an analytical description of the dynamics of the three-qubit Dicke
model using the adiabatic approximation in the parameter regime where the
qubits transition frequencies are far off-resonance with the field frequency
and the interaction strengths reach the ultra-strong coupling regimes.
Qualitative differences arise when comparing to the single- and two-qubit
systems - simple analytic formulas show that three revival sequences produce
a three-frequency beat note in the time evolution of the population. We find
an explicit way to estimate the dynamics for the qubit-field and qubit-qubit
entanglement inside the three-qubit system in the ultra strong coupling regime,
and the resistance to the sudden death proves the robustness of the GHZ state.
\end{abstract}

\maketitle

\section{Introduction}

Recent experimental studies have shown that the ultrastrong coupling
regime, where the coupling strength is some tenths of the mode frequency,
can be achieved in a number of implementations such as
superconducting circuits \cite{Niemczyk, Forn-Diaz, Fedorov, Bourassa},
semiconductor quantum wells \cite{Gunter, Todorov, Anappara}, possibly also in
surface acoustic waves \cite{Gustafsson} and trapped ions \cite{Pedernales}.
The fast-growing interest in the ultrastrong coupling regime is motivated
not only by theoretical predictions of novel fundamental properties \cite%
{Liberato, Felicetti2, Felicetti3} but also by potential applications in
quantum computing tasks \cite{Nataf, Kyaw}. The advent of these impressive
experimental results prompts a number of theoretical efforts to give
analytical solutions for the quantum Rabi and Dicke models \cite{Rabi, Dicke}
by applying various techniques \cite{Liu, Braak, Braak1, Chen1, Chen5}. On
the other hand, the models are expanded to more general cases, including
different qubits \cite{Peng, Peng1, Chilingaryan, Mao, Chen6}, anisotropic
couplings \cite{Xie, Cui, Zhong}, a finite-size ensemble of interacting
qubits \cite{Robles} and two-photon interactions \cite{Felicetti}, to name only a few.

In particular, people start to tackle the entanglement features both between the qubit
and the field and inside the qubit system, yet with the rotating wave approximation
\cite{Tessier, Youssef, Obada, Lopez}.
Entanglement, as a fundamental quantum mechanical tool describing the
non-local correlations between quantum objects, lies at the heart of
quantum information sciences \cite{Preskill, Islam, Hensen}. It is highly expected that
nontrivial population and entanglement dynamics would emerge in the ultrastrong
coupling regime where the RWA fails. For the Rabi model, a displaced Fock state method
\cite{Irish1} is developed to analytically predict the time evolution of the qubits occupation
probability in the case of strong coupling and large detuning.
The key step is the adiabatic approximation which nicely truncates the
system Hamiltonian into a block diagonal form and the resulted solutions are utilized to study the
entanglement dynamics in the two-qubit system \cite{Agarwal, Mao}.
Specifically, a simple expression of the concurrence \cite{Wootters} for the two qubits
is given analytically and the entanglement sudden death appears even in the inhomogeneous
coupling case. The circuit quantum electrodynamics (QED) architecture offers considerable
potential for simulating such dynamics following an analog-digital approach \cite{Mezzacapo}.

The motivation of this study is stimulated by a lack of an analytical analysis of the bipartite
entanglement inside the multi-qubit system in the ultras-trong coupling regime. Typical model
for this is the three-qubit Dicke model characterized by a realistic realization of a GHZ state
\cite{Greenberger}. Recently, it has been shown that superconducting circuit technology
allows to exploit the the dynamical Casimir effect physics as a useful resource for
the generation of highly entangled states for multi superconducting qubits \cite{Felicetti4}.
It is thus desirable to demonstrate whether the sudden death of entanglement
would survive the dissipative effects for the strong and
ultrastrong coupling regimes \cite{Rossatto}. Here, we show the
robustness of three-qubit GHZ state against the
interaction and the energy exchange between the qubits and the field in the
Dicke model, which could be of importance for future applications e.g. in
quantum cryptography \cite{Hao, Kempe}, quantum computation \cite{Pedersen}
and quantum gates \cite{Monz1, Joshi, Ivanov}.

This paper is organized as follows. We solve the three-qubit Dicke model in a
spin-3/2 subspace and derive the analytical eigen solutions by means of the
adiabatic approximation in Sec. II. These results are applied to study the population
dynamics of the three qubits coupled to, respectively, a Fock state and a coherent
state of the oscillator in Sec. III. The spectrum of the multi-revival signal is analyzed and
compared to the numerical calculation without the adiabatic approximation. Then,
we explore the entanglement dynamics for three
qubits starting from the GHZ state and the field in a coherent state and show the robustness
of the GHZ state through the bipartite entanglement measure $I$ tangle in Sec. V.
Finally, a brief summary is presented in Sec. VI.

\section{Eigensolution of the three-qubit Dicke model}

We consider the three-qubit Dicke model described by the following
Hamiltonian ($\hbar =1$) \cite{Dicke,Chen1,Braak1}%
\begin{equation}
H=\omega _{c}a^{\dagger }a-\omega J_{x}+2g\left( a^{\dagger }+a\right) J_{z}.
\label{H}
\end{equation}%
Here $a^{\dagger }$ ($a$) is the creation (annihilation) operator of the
single bosonic mode with frequency $\omega _{c}$, $\omega $ denotes the
qubit splitting, the constant $g$ represents the coupling between the qubit
and field mode, and the total spin operator is the sum of the Pauli
operators of the individual qubits, i.e. $J_{i}=\sum_{\alpha =1}^{3}\sigma
_{i}^{\alpha }/2$ ($i=x,y,z$).
Note that $J^{2}$ commutes with the Hamiltonian (\ref{H}), i.e. $\left[
J^{2},H\right] =0$, this provides a splitting of the eight-dimensional spin
space into a quadruplet state space $j=3/2$ and two doublet state spaces $%
j=1/2$,
\begin{equation}
H=H^{3/2}\oplus H^{1/2}\oplus H^{1/2}.
\end{equation}%
which comes from three possible standard Young tableaux ${\scriptsize \young%
(123)}$, ${\scriptsize \young(12,3)}$, and ${\scriptsize \young(13,2)}$ in
representation theory of permutation group theory \cite{Dukalski}. The
three-qubit Dicke model thus decomposes into a system of one spin-3/2 and
two spin-1/2 Rabi models \cite{Braak1}. While both models have been solved
using the displaced Fock space method \cite{Irish1,Liu,Chen1} and
Bargmann-space techniques \cite{Braak,Braak1}, little attention has been
paid to the analytical dynamics of the qubit occupation probability due to
the cumbersome task in extracting the analytical solution in both
formulations. Indeed the power series must be terminated in the
transcendental function $G_\pm(x)$ \cite{Braak}, or the expansion of the
wave function in terms of displaced Fock space should be truncated in a
finite $N_{tr}$ subspace \cite{Liu,Chen1}.

Here, similar to the cases of single- \cite{Liu,Irish1} and two-qubit \cite%
{Agarwal, Mao} Rabi model, we apply the adiabatic approximation to the
numerical solutions when the frequencies of the qubits are much smaller than
the oscillator frequency $\omega \ll \omega_{c}$. In this displaced
oscillator basis the Hamiltonian may be truncated to a block-diagonal form
and the blocks solved individually. We shall confine ourselves in the
following to the system with four-dimensional spin-subspace of $j=3/2$, due
to the fact that all interesting dynamics in the three-qubit system prepared
in the experiments is confined in this subspace, and best of all, this method is the
most effective way to study analytically the dynamical properties of
three qubits. The Hamiltonian $H^{3/2}$ reduces to $4\times 4$ block
diagonal form, i.e. $H^{3/2}=\sum_{n=0}^{\infty }\oplus H_{n}$ with
\begin{equation}
H_{n} = \left(
\begin{matrix}
\epsilon^n_{3/2} & \sqrt{3}\Omega _{n} & 0 & 0 \\
\sqrt{3}\Omega _{n} & \epsilon^n_{1/2} & 2\Omega _{n} & 0 \\
0 & 2\Omega _{n} & \epsilon^n_{1/2} & \sqrt{3}\Omega _{n} \\
0 & 0 & \sqrt{3}\Omega _{n} & \epsilon^n_{3/2}%
\end{matrix}%
\right).  \label{H2}
\end{equation}%
where $\epsilon^n_{m}=\omega _{c}\left( n-\beta _{m}^{2}\right)$. The system
is spanned by the the joint spin-field space $\left\vert 3/2,m\right\rangle
\left\vert n\right\rangle _{A_{m}}$ where $\left\vert j,m\right\rangle $ are
eigenstates of $J^{2}$ and $J_{z}$ with eigenvalues $j(j+1)$ and $%
m=-j,-j+1,...,j$, respectively, and the displaced Fock states satisfy $%
A_{m}^{\dagger }A_{m}\left\vert n\right\rangle _{A_{m}}=n\left\vert
n\right\rangle _{A_{m}}$with $A_{m}=a+\beta _{m}$ and $\beta _{m}=m\alpha$
and $\alpha=2g/\omega _{c}$. The off-diagonal elements in the matrix are
given by%
\begin{equation}
\Omega _{n}=-\frac{\omega }{2}e^{-\frac{\alpha^{2}}{2}}\sum_{l=0}^{n}\frac{%
(-1)^{n-l}n!}{l![(n-l)!]^{2}}\alpha ^{2\left( n-l\right) }.  \label{omega}
\end{equation}%
It can be easily proved that the parity operator defined in the spin-3/2
subspace as $\Pi ^{3/2}=\exp \left[ i\pi \left( 3/2-J_{x}+a^{\dagger
}a\right) \right] $ with eigenvalues $\kappa =\pm 1$, commutes with the
Hamiltonian, i.e. $\left[ H^{3/2},\Pi ^{3/2}\right] =0$. Accordingly, the
Hilbert space splits into two mutually orthogonal subspaces with even and
odd parities \cite{Braak1}. The four eigenstates of $H_{n}$ are
non-degenerate and can be classified uniquely by one quantum number.
However, for numerical solutions beyond the adiabatic approximation, it is
necessary to use the parity operator to classify the degeneracies that take
place between levels of states with different parity. For consistency, we
use parity invariance $\left[ H_n,\Pi_n\right] =0$ with $\Pi_n$ an
anti-diagonal matrix $(-1)^n \text{adiag}[1,1,1,1]$, to further block
diagonalize $H_{n}$ as $H_{n}=\sum_{\kappa=\pm 1 }\oplus H_{n}^{\kappa }$
with
\begin{equation}
H_{n}^{\kappa }=\left(
\begin{array}{cc}
\epsilon^n_{3/2} & \sqrt{3}\Omega _{n} \\
\sqrt{3}\Omega _{n} & \epsilon^n_{1/2} +2\xi \Omega _{n}%
\end{array}%
\right)  \label{H3}
\end{equation}%
and $\xi =\kappa \left( -1\right) ^{n}$. The energy levels are given by%
\begin{equation}
E_{n}^{\kappa \pm } = n\omega_c+\xi \Omega _{n} - 5g^2/\omega_c \pm \theta
_{n}^{\kappa }  \label{eigenergy1}
\end{equation}
with $\theta _{n}^{\kappa }=\sqrt{( \xi \Omega _{n}+4g^2/
\omega_c)^2+3\Omega _{n}^{2}}$. The corresponding eigenstates in the
spin-field space are
\begin{equation}
\left\vert \psi _{n}^{\kappa \pm }\right\rangle =d_n^{\kappa \pm}\left(
c_{n}^{\kappa \pm },1,\xi ,\xi c_{n}^{\kappa \pm }\right) ^{T}
\label{eigenstate1}
\end{equation}%
with $c_{n}^{\kappa \pm }=\sqrt{3}\Omega _{n}/\left( \xi \Omega _{n}+4g^2/
\omega_c\pm \theta _{n}^{\kappa }\right)$ and $d_n^{\kappa \pm}=1/\sqrt{%
2\left\vert c_{n}^{\kappa \pm}\right\vert^2+2} $. We restrict the analysis
in the following to the case of $\left\vert \Omega _{n}\right\vert \gg 4g^2/
\omega_c$ fulfilled by most experimental systems in the ultra-strong
coupling regime $g \leq 0.08 \omega_c$, which enables us to achieve an
analytical dynamics below. The eigenvalues are therefore simplified to%
\begin{equation}
E_{n}^{\kappa \pm} = n\omega _{c}+(\xi \mp 2) \Omega _{n},
\label{eigenergy2}
\end{equation}
and the corresponding eigenfunctions are%
\begin{equation}
\left\vert \psi _{n}^{\kappa \pm}\right\rangle =\sqrt{\frac{2\mp \xi}{8}}%
\left( \frac{\sqrt{3}}{\xi\mp 2},1,\xi ,\frac{\xi \sqrt{3}}{\xi\mp 2}\right)
^{T}.  \label{eigenstate3}
\end{equation}%
We observe that the parity $\kappa $ and the photon number $n$ in the
displaced Fock state are independent to each other. This is due to the fact
that in constructing the unitary transformation which brings the Hamiltonian
into a block diagonal form (\ref{H3}), a phase difference $n\pi $ is
introduced in the superposition of two displaced Fock states $\left\vert
n\right\rangle _{A_{m}}$ in opposite directions so that symmetric and
antisymmetry superposition states of the corresponding bases are
respectively even and odd parities. This situation resembles the parity of
the ground state and the first excited state in the standard quantum
tunneling model of double well potential.

\section{Population dynamics}

After a detailed discussion of the energy spectrum, we now turn to the
study of population dynamics of the qubits. In particular, we will examine
the dynamics with all three qubits being excited to the upper level $%
\left
\vert eee\right \rangle$, while the initial state of the oscillator
is prepared in the displaced Fock basis corresponding to it, i.e. $%
\left\vert \Psi \left( 0\right) \right\rangle =\left\vert
3/2,3/2\right\rangle \left\vert n\right\rangle _{A_{3/2}}$, which is
expressed as the linear combination of the eigenvectors (\ref{eigenstate1})
\begin{equation}
\left\vert \Psi \left( 0\right) \right\rangle =\sum\limits_{\kappa ,\tau
}d_{n}^{\kappa \tau } c_{n}^{\kappa \tau } \left\vert \psi _{n}^{\kappa \tau
}\right\rangle.
\end{equation}%
Then at subsequent time $t$ the probability to find three qubits in the
initial state $\left\vert 3/2,3/2\right\rangle $ is easily obtained
\begin{equation}
P_{1}\left( n,t\right) =\left\vert _{A_{3/2}}\left\langle n\right\vert
\left\langle 3/2,3/2|\Psi \left( t\right) \right\rangle \right\vert ^{2}.
\label{FP}
\end{equation}%
Substituting the simplified eigensolutions (\ref{eigenergy2}) and (\ref%
{eigenstate3}) into Eq. (\ref{FP}), we find that the probability is composed
of three oscillating frequencies
\begin{eqnarray}
P_{1}\left( n,t\right) &=&\frac{1}{32}\left( 10+15\cos \left( 2\Omega
_{n}t\right) \right.  \notag \\
&&\left. +6\cos \left( 4\Omega _{n}t\right) +\cos \left( 6\Omega
_{n}t\right) \right),
\end{eqnarray}%
while in the single- and two-qubit Rabi models we have respectively one and
two frequencies dominating the evolution.

\begin{figure}[t]
\includegraphics[width=0.45\textwidth]{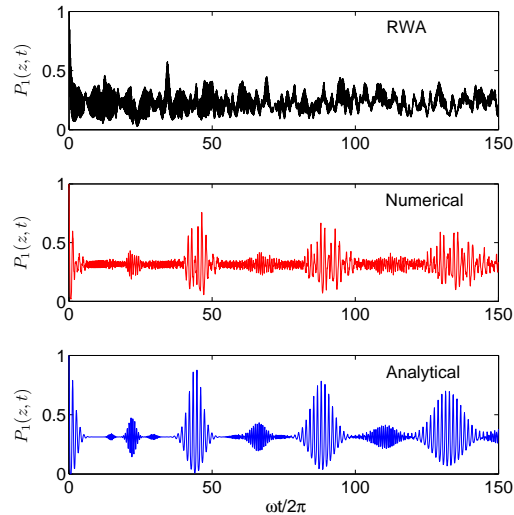}
\caption{(Color Online) Probability $P_{1}(z,t)$ of finding three qubits in
the initial state $\left\vert 3/2,3/2\right\rangle $ as a function of $%
\protect\omega t/2\protect\pi $ for $\protect\omega =0.15\protect\omega _{c}$%
, $g=0.08\protect\omega _{c}$ and $z=3$. Note the breakup in the main
revival peaks of the numerical evaluation, which comes from the $\protect%
\omega-2\protect\omega-3\protect\omega$ beat note, is not included in the
analytic calculation.}
\label{Fig1}
\end{figure}

If instead, initially the oscillator is displaced from a coherent state $%
\left\vert z \right \rangle$, i.e.
\begin{equation}
\left\vert \Psi \left( 0\right) \right\rangle =\sum\limits_{n=0}^{+\infty }%
\frac{e^{-|z|^{2}/2}z^{n}}{\sqrt{n!}}\left\vert 3/2,3/2\right\rangle
\left\vert n\right\rangle _{A_{3/2}},  \label{initial state}
\end{equation}%
which is the closest quantum state to a classical wave and more realistic
for describing the oscillator, the probability of three qubits remaining in
their initial state $\left\vert 3/2,3/2\right\rangle $ is calculated by
tracing over all Fock states as
\begin{eqnarray}
P_{1}\left( z,t\right) &=&\left\langle 3/2,3/2|Tr_{F}\rho
(z,t)|3/2,3/2\right\rangle  \notag \\
&=&\sum\limits_{n=0}^{+\infty }p\left( n\right) P_{1}\left( n,t\right) .
\label{population}
\end{eqnarray}%
Here $\rho (z,t)=|\Psi (t)\rangle \langle \Psi (t)|$ is the density matrix
of the system and the normalized Poisson distribution is defined as $p\left(
n\right) =e^{-\left\vert z\right\vert ^{2}}\left\vert z\right\vert ^{2n}/n!$%
. Following the procedure established previously for two-qubit model \cite%
{Agarwal} by keeping only three terms $l=n,n-1,n-2$ in the summation of $%
\Omega _{n}$ and replacing the Poisson distribution by a Gaussian one for
big enough $|z|$, we may reduce Eq. (\ref{population}) into the following
analytical form
\begin{equation}
P_{1}\left( z,t\right) =\frac{1}{32}\text{Re}\left[ 10+15S\left( t,\omega
\right) +6S\left( t,2\omega \right) +S\left( t,3\omega \right) \right] ,
\label{population3}
\end{equation}%
with $S\left( t,\omega \right) =\sum\nolimits_{k=0}^{+\infty }S_{k}\left(
t,\omega \right) $. The collapse and revival of the probability $P_{1}\left(
z,t\right) $, which is approximated with fairly good accuracy by the
sufficiently simple function $S_{k}\left( t,\omega \right) $, is obvious
here and the individual revival function
\begin{equation}
S_{k}\left( t,\omega \right) =h_{k}\exp \left( \Phi _{Re}+i\Phi _{Im}\right)
\label{sk}
\end{equation}%
with a height $h_{k}=\left( 1+\pi ^{2}k^{2}f^{2}\right) ^{-1/4}$ and
\begin{eqnarray}
\Phi _{Re} &=&-\frac{1}{2}h_{k}^{4}\left( \mu -\mu _{k}\right) ^{2}f\alpha
^{2}, \\
\Phi _{Im} &=&\frac{1}{2}\tan ^{-1}\left( \pi kf\right) +\mu (1-f) +2\pi k
|z|^2,
\end{eqnarray}%
describes the evolution around the $k$-th revival time $t_{k}^{rev}=\mu
_{k}/\omega$ where we have defined $f=\left\vert \alpha z\right\vert ^{2}$, $%
\mu =\omega te^{-\alpha ^{2}/2} $, and $\mu _{k}=\pi k\left( f+2\right)
/\alpha ^{2}$. During each peroid $\Delta t=\pi \left( f+2\right)
/\omega\alpha ^{2}$, however, the signals in $S_{k}\left( t,2\omega \right)$
and $S_{k}\left( t,3\omega \right)$ revive twice and three times
respectively. We thus get three revival sequences in the evolution of the
probability $P_{1}\left( z,t\right) $. The envelope and the fast oscillatory
of the revival signal are determined by $\Phi _{Re}$ and $\Phi _{Im}$
respectively \cite{Agarwal}.

In Fig. \ref{Fig1}, a comparison of the analytic formula derived for $%
P_{1}\left( z,t\right) $ and the numerical calculations is made in the
parameter regime where the coupling strength is strong enough to invalidate
the RWA. We see that with the time increasing the equilibrium value $10/32$,
about which the revival signal oscillates, is smaller than $16/32$ in the
single qubit model and $12/32$ in the two-qubit model \cite{Agarwal} due to
the involvement of higher order harmonic signals in the probability. The
width of the successive revival signals keeps increasing as $\delta \mu_k=%
\sqrt{1+\pi^2 k^2 f^2}/|z|\alpha^2$ that leads to the mergence of the third
harmonic signal into the first and second ones after several revival
periods. The salient feature of the three-qubit model as demonstrated above
is that the revival signals corresponding to the three oscillating terms $%
S\left( t,\omega \right), S\left( t,2\omega \right)$ and $S\left( t,3\omega
\right)$ produce a beat note of $\omega-2\omega-3\omega$. The three revival
sequences in the evolution of $P_{1}\left( z,t\right) $ are even clearer in
the Fourier analysis $\bar{P}_1(z,\nu)$ defined as
\begin{equation}
\bar{P}_1(z,\nu)=\int_0^{+\infty} dt P(z,t) e^{-i2\pi \nu t}
\end{equation}
which is presented in Fig. 2. The spectral signals $\bar{P}_1(n,\nu)$
corresponding to the probability of displaced Fock state are $\delta$
functions located at $2\Omega_n$, $4\Omega_n$ and $6\Omega_n$, respectively.
The involvement of Fock states of many photons in the coherent state leads
to a broad distribution of the spectral functions for $P_1(z,t)$ at a
fundamental frequency $\omega^*=\omega e^{-\alpha^2 /2}(1-|z|^2\alpha^2)$,
which is $0.76 \omega$ for $g=0.08 \omega_c$ and $z=3$, as well as at the
second and third harmonics with decreasing magnitude. This contrasts the
single and double revival sequences for the single and two-qubit systems as
a consequence of having only one and two Rabi frequencies, respectively,
which have been showed in \cite{Agarwal}. The analytical results reproduce
the multiple revival sequences for the three-qubit model, except that
breakups appear in the main and the second harmonic revival frequencies if
no adiabatic approximation is made, which can be compared with the RWA case
in \cite{Deng}. We find that the RWA completely breaks down in the ultra
strong coupling parameter regime considered here.

\begin{figure}[t]
\includegraphics[width=0.45\textwidth]{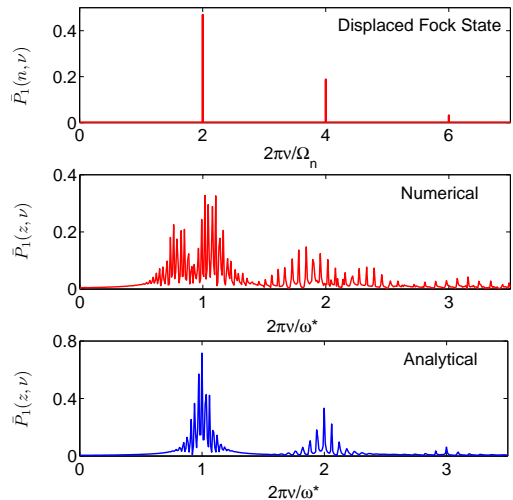}
\caption{(Color Online) The Fourier analysis of the probability revival for
the displaced Fock state (top) and displaced coherent state (middle and
bottom panels for numerical and analytical calculation respectively) of the
oscillator with all three qubits being excited to the upper level. Three
revival sequences in the dynamics produce a beat note of $\protect\omega-2%
\protect\omega-3\protect\omega$ and the breakups in the fundamental and the
second harmonic frequencies without the adiabatic approximation. The
corresponding parameters are the same as in Figure \protect\ref{Fig1}.}
\label{Fig2}
\end{figure}

\section{Entanglement behaviors}

The entanglement properties of two identical qubits strongly coupled to a
single-mode radiation field have recently been studied in \cite{Agarwal,
Chen}, where the entanglement sudden death does appear in the numerical and
analytic calculations. However, qualitative differences should arise in the
case of the three-qubit system. It is widely accepted nowadays that
entangled states of multi-particle systems are the most promising resource
for quantum information processing \cite{Bennett, Nielsen, H1}. Thus, it is
highly desirable to explore the entanglement dynamics of the three-qubit
Dicke model.

In this section we provide an easily computable formula for the entanglement
dynamics when the field is initially in a coherent state $\left\vert z
\right\rangle$ and the three qubits are initially in the form of a familiar
GHZ state $\frac{1}{\sqrt{2}}\left( \left\vert eee \right\rangle+\left\vert
ggg \right\rangle \right)$, i.e.
\begin{equation}
\left\vert \Psi \left( 0\right) \right\rangle =\frac{1}{\sqrt{2}}\left(
\left\vert 3/2,3/2\right\rangle +\left\vert 3/2,-3/2\right\rangle \right)
\left\vert z\right\rangle .  \label{intial state1}
\end{equation}%
For small values of $\alpha$, we may expand the state $\left\vert
n\right\rangle $ in the displaced Fock space and the most important
contribution in the summation over $n$ comes from the terms with the same $n$%
, which is equivalent to take $\left\vert n\right\rangle \approx \left\vert
n\right\rangle _{A_{m}}$ \cite{Agarwal}. This approximation
gives the state at subsequent time $t$ as
\begin{equation}
\left\vert \Psi \left( t\right) \right\rangle =\sum\limits_{n,\kappa \tau }%
\sqrt{\frac{p\left( n\right)}{2}}\left( 1+\xi \right) \!{d_{n}^{\kappa \tau
}c_{n}^{\kappa \tau }}\left\vert \psi _{n}^{\kappa \tau }\right\rangle
e^{-iE_{n}^{\kappa \tau }t}.  \label{time state3}
\end{equation}%
To examine the entanglement evolution of the system we calculate the reduced
density matrix of the qubits by tracing over the quantum field
\begin{equation}
{\rho }_{Q}\left( t\right) =\sum\limits_{n}\left\langle n|\Psi \left(
t\right) \right\rangle \left\langle \Psi \left( t\right) |n\right\rangle,
\end{equation}%
which can be reduced to the following matrix form%
\begin{equation}
{\rho }_{Q}\left( t\right) =\left(
\begin{array}{cccc}
\frac{1}{4} & 0 & \frac{\sqrt{3}S\left( t,2\omega \right) }{4} & 0 \\
0 & 0 & 0 & 0 \\
\frac{\sqrt{3}S^{\ast }\left( t,2\omega \right) }{4} & 0 & \frac{3}{4} & 0
\\
0 & 0 & 0 & 0%
\end{array}%
\right)  \label{density}
\end{equation}%
in the eigen basis $\left\vert 3/2,m\right\rangle _{x}$ of spin $J_{x}.$
\begin{figure}[t]
\includegraphics[width=0.45\textwidth]{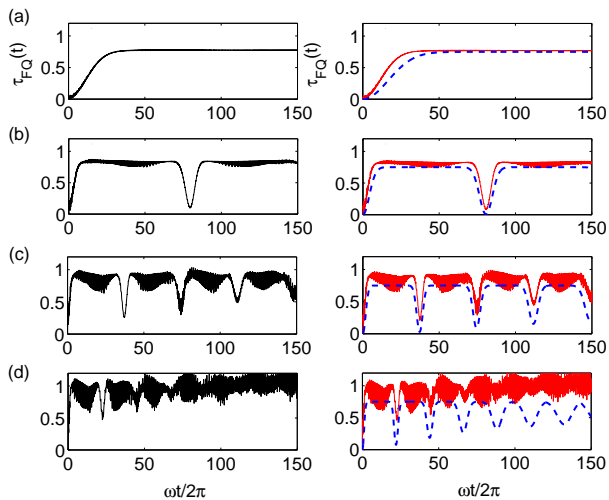}
\caption{(Color Online) The $I$ tangle between the qubits and light field as
a function of $\protect\omega t/2\protect\pi $ for transition frequency $%
\protect\omega =0.15\protect\omega _{c}$ and different coupling strengthes $%
g=0.02\protect\omega _{c}(a),0.04\protect\omega _{c}(b),0.06\protect\omega %
_{c}(c)$ and $0.08\protect\omega _{c}(d)$, given by numerical method (left),
the adiabatic approximation (red solid line, right) and analytical approach
(blue dashed line, right).}
\label{Fig3}
\end{figure}

The entanglement between the field and the qubits may be described by the $I$
tangle
\begin{equation}
\tau _{FQ}\left( t\right) =2\left( 1-tr\left( \rho _{Q}^{2}\right) \right) ,
\label{tangle3}
\end{equation}%
which is introduced in \cite{Rungta} and applicable to infinite-dimensional
bipartite systems \cite{Youssef}. It runs from zero for a product state to
the maximum value $2\left( d-1\right) /d=1.75$ with $d=\min \left(
d_{1,}d_{2}\right) $ for a maximally entangled state, where $d_{1},d_{2}$
are respectively the dimensions of the three-qubit system and the photon
field. The analytic expression for the reduced density matrix (\ref{density}%
) allows us to obtain the explicit formula
\begin{equation}
\tau _{FQ}\left( t\right) =\frac{3-3\left\vert S\left( t,2\omega \right)
\right\vert ^{2}}{4}.  \label{tangle2}
\end{equation}%
In Fig. \ref{Fig3} we plot the time evolution of $I$ tangle $\tau _{FQ}$ for
various values of $g$. Should we adopt the expression (\ref{eigenstate1})
for the eigenstates $\left\vert \psi _{n}^{\kappa \tau }\right\rangle $ and
do not take into account $\left\vert n\right\rangle \approx \left\vert
n\right\rangle _{A_{m}}$, the adiabatic approximation produces a quite
accurate result in the ultrastrong coupling regime as shown in the right
column of Fig. \ref{Fig3}. The analytic result determined by formula (\ref%
{tangle2}) agrees well with the envelope of the numerically evaluated
result, but fails in describing the long time behavior when the coupling
strengths are sufficiently large which is evident in Fig. \ref{Fig3}. The $I$
tangle $\tau _{FQ}$ starts from zero for the initial product state (\ref%
{intial state1}), and undergoes periodic weakening and recovery with the
oscillation period getting smaller and smaller for increasing coupling
strengths, which however could never reach the maximum entanglement value of
the system. We see that the field-qubit entanglement exhibits collapse and
revival and the analytic formula predicts correctly the main features of the
individual entanglement revival signals.

\begin{figure}[t]
\includegraphics[width=0.45\textwidth]{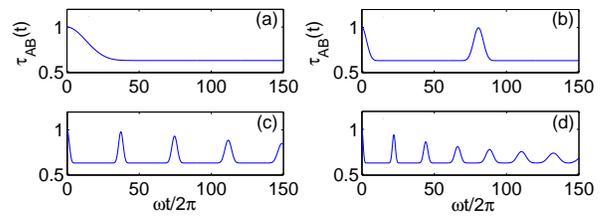}
\caption{(Color Online) The $I$ tangle between the three qubits as a
function of $\protect\omega t/2\protect\pi $ given by the analytical
calcultion. The corresponding parameters are the same as in Figure \protect
\ref{Fig3}.}
\label{Fig4}
\end{figure}

The initially pure state of the qubits evolves into a mixed state described
by the reduced density matrix $\rho _{Q}$ in Eq. (\ref{density}). We may
analytically study the entanglement between the qubits in the following. The
measures of entanglement for mixed states depend on the pure state
decompositions, in this way the main difficulty is to find the minimization
over all decompositions of mixed state into pure states. However, our
analytic method provides a particular case, where $\rho _{Q}$ is a rank-2
mixed state of a qubit and a qudit in the basis of the three-qubit product
states. Thus we may discuss the properties of the entanglement of three
qubits system using the $I$ tangle proposed by Osborne et al. \cite{Osborne}%
. As a good mixed-state entanglement measure for three qubits, the $I$
tangle $\tau _{AB}\left( t\right) $ between one qubit (subsystem A) and the
other two qubits (subsystem B) is given by the formula \cite{Osborne}
\begin{equation}
\tau _{AB}\left( t\right) =Tr\left( \rho _{Q}\tilde{\rho}_{Q}\right)
+2\lambda _{\min }\left[ 1-Tr\left( \rho _{Q}^{2}\right) \right] ,
\label{tangle}
\end{equation}%
where the universal state inverter is defined as $\tilde{\rho}%
_{Q}=I_{A}\otimes I_{B}-\rho _{A}\otimes I_{B}-I\otimes \rho _{B}+\rho _{Q}$
with $\rho _{A}=Tr_{B}\left( \rho _{Q}\right) $ and $\rho _{B}=Tr_{A}\left(
\rho _{Q}\right)$ and $\lambda _{\min }$ is the smallest eigenvalue of a
real symmetric $3\times 3$ matrix $M$ as defined in \cite{Osborne, Liberti}.
A tedious yet straightforward calculation gives
\begin{equation}
M =\frac{1}{3(1+3|S|^2)}\left(
\begin{array}{ccc}
2+2|S|^2 & 0 & \frac{4|S|}{\sqrt{3}} \\
0 & \frac{-1-|S|^2}{2} & 0 \\
\frac{4|S|}{\sqrt{3}} & 0 & \frac{-1+9|S|^2}{3}%
\end{array}%
\right)  \label{M}
\end{equation}
and $\lambda_{min}=-(1+|S|^2)/(6+18|S|^2)$. Inserting this and the analytic
expression (\ref{density}) into Eq. (\ref{tangle}) gives a very simple
result for the bipartite entanglement
\begin{equation}
\tau _{AB}\left( t\right) =\frac{5+20\left\vert S\left( t,2\omega \right)
\right\vert ^{2}+7\left\vert S\left( t,2\omega \right) \right\vert ^{4} }{%
8\left( 1+3\left\vert S\left( t,2\omega \right) \right\vert ^{2}\right) }.
\label{tangle1}
\end{equation}%
The evolution of the $I$ tangle for three qubits is plotted in Fig. \ref%
{Fig4}. In contrast to the field--qubit entanglement $\tau _{FQ}\left(
t\right)$, it starts from the maximum value and then decreases to a finite
steady value where entanglement sudden death is absent. When the
field--qubit entanglement becomes rather weak this qubit-qubit entanglement
increases rapidly to a large value, as can be seen in Fig. \ref{Fig3} and
Fig. \ref{Fig4}. It turns out that the decrease of $\tau _{FQ}\left(
t\right) $ is directly related to the growth of $\tau _{AB}\left( t\right) $
in the form of Eq. (\ref{tangle2}) and Eq. (\ref{tangle1}), which predict
correctly the time, height, and width of the individual entanglement
oscillation. This indicates that the initial entanglement between the qubits
withstands the interaction and the energy exchange between the qubits and
the field. In this way we provide for the first time an explicit analytical
formula for the robustness of the GHZ state in the three-qubit Dicke model.

\section{Conclusion}

In conclusion, we have analyzed the population and entanglement dynamics of
three qubits within the adiabatic approximation. It works very well in the
ultrastrong coupling regime under the assumption that the qubit frequencies
are much smaller than the field frequency. The remarkable feature of
population dynamics in the three-qubit model is that the three revival
sequences in the evolution of the probability produce a three-frequency beat
note. Moreover, the analytic formulas of the $I$ tangle for the pure state
of field-qubit system and mixed state of three-qubit exhibit their
excellence in entanglement characterization and distribution. This is the
first to present the robustness of the GHZ state in the form of the analytic
expressions in the three-qubit Dicke model. The sudden death of the
entanglement is avoided in the three-qubit system, which are
qualitatively different from the two-qubit case studied in \cite{Agarwal,
Mao, Chen}. A practically relevant application of our result lies in the
quantum information process with circuit QED, where three-qubit entangled
states are involved.

\section*{Acknowledgements}

This work is supported by NSF of China under Grant Nos. 11234008 and
11474189, the National Basic Research Program of China (973 Program) under
Grant No. 2011CB921601, Program for Changjiang Scholars and Innovative
Research Team in University (PCSIRT)(No. IRT13076).

\end{document}